# Backcalculation of the disease-age specific frequency of secondary transmission of primary pneumonic plague

**Running title**

Infectiousness of pneumonic plague


**Author:**

Hiroshi Nishiura

**Affiliation:**

Theoretical Epidemiology, University of Utrecht, Yalelaan 7, 3584 CL, Utrecht, The

Netherlands




# Abstract


**Aim**     To assess the frequency of secondary transmissions of primary pneumonic plague relative to the onset of fever.

**Methods**     A simple backcalculation method was employed to estimate the frequency of secondary transmissions relative to disease-age. A likelihood-based procedure was taken using observed distributions of the serial interval (n = 177) and incubation period (n = 126). Furthermore, an extended model was developed to account for the survival probability of cases.

**Results**     The simple backcalculation suggested that 31.0% (95% confidence intervals (CI): 11.6, 50.4) and 28.0 % (95% CI: 10.2, 45.8) of the total number of secondary transmissions had occurred at second and third days of the disease, respectively, and more than four-fifths of the secondary transmission occurred before the end of third day of disease. The survivorship-adjusted frequency of secondary transmissions was obtained, demonstrating that the infectiousness in later stages of illness was not insignificant and indicates that the obtained frequencies were likely biased on underlying factors including isolation measures.

**Conclusion**     The simple exercise suggests a need to implement countermeasures during pre-clinical stage or immediately after onset. Further information is needed to elucidate the finer details of the disease-age specific infectiousness.


# Keywords



- 2 -



## Correspondence

3     Tel: +31 30 253 1233; Fax: +31 30 252 1887

4     E-mail: h.nishiura@uu.nl





## INTRODUCTION

Primary pneumonic plague is contracted when the causative agent, *Yersinia pestis*, a category A agent [1], is inhaled during human-to-human transmission, and is considered one of the diseases most likely to be caused in the event of a bioterrorist attack. The case–fatality reaches almost 100% without appropriate chemoprophylaxis or treatment immediately after onset [2]. Despite studies exploring other forms of the infection (e.g., bubonic plague [3]), the transmission mechanisms of primary pneumonic plague have remained unclear mainly due to limited data availability. To date, mathematical studies assuming population dynamics have suggested that rapid countermeasures (e.g., chemoprophylaxis and contact tracing) are crucial to contain the outbreak [4,5], and in light of this, it is important to further understand the natural history of this disease.

In particular, infectiousness relative to the time-course of disease (i.e., disease-age) plays a key role in determining the feasibility of disease control [6]. Targeted control measures, including isolation and contact tracing, have to be implemented as early as possible during the infectious period, and thus, their effectiveness largely depends on the time-course of infectiousness. One approach is to quantify how the pathogen load changes over time by using the most sensitive microbiological techniques (e.g., Real-Time polymerase chain reaction), but such observations are practically limited to the



time after onset of symptoms and the pathogen load information can only be a useful

measure of infectiousness if it is correlated with actual transmission. Although previous

studies tended to focus on the overall transmission potential measured by the basic

reproduction number, $R_0$, the average number of secondary cases arising from a single

primary case in a fully susceptible population [7], disease-age specific infectiousness

was, for simplicity, frequently assumed constant during the infectious period and almost

ignored for diseases with acute course of illness. To improve this understanding, the

present study proposes an epidemiological evaluation method, based on the distribution

of the incubation period and the transmission network (who acquired infection from

whom), to estimate how plague infectiousness varies over the course of illness. The

present study was aimed at estimating the frequency of secondary transmissions of

pneumonic plague relative to disease-age using historical outbreak data.

## MATERIALS AND METHODS

### Theoretical basis and data

First, this study applied a simple backcalculation method to estimate the frequency

of secondary transmissions relative to disease-age, as recently employed for a study of

smallpox [8], and second, attempts to extend this method. In the following, "disease-



1   age" is measured as the time since onset of symptoms (i.e., disease-age $t = 0$ denotes the

2   onset of fever). This approach is based on two of the known distributions, the serial

3   interval and incubation period. Serial interval is defined as the time from symptom

4   onset in a primary case to symptom onset in a secondary case [9,10]. The distribution of

5   serial intervals can be extracted from the transmission network in historical datasets,

6   which includes information on who infected whom [11,12]. This study uses a total of

7   177 serial intervals from 4 outbreaks; 88 intervals in Manchuria (1910-11), 32 in

8   Mukden (1946), 17 in NW Madagascar (1957), and 40 in Central Madagascar (1997)

9   [13-16]. The mean (median and standard deviation (SD)) of the intervals was 5.4 (5.0

10  and 3.0) days. The minimum and maximum intervals were 1 and 18 days, respectively. I

11  denote the number of observed serial intervals of length $t$ days by $s_t$. The incubation

12  period is the time from infection to onset of disease (i.e., fever). The distribution was

13  obtained from a study of Kasai [17] who determined the time of exposure from contact

14  tracing information during the largest outbreak in Manchuria from 1910–11 (n = 126;

15  Figure 1a), with a mean of 4.5 (5.0, 1.3) days. Since the distribution did not reasonably

16  fit standard statistical distributions, and because the following model is constructed in

17  discrete time (i.e., the data in daily precision), I used the normalized frequency (i.e.,



1    observed numbers for each day divided by sample size) as the incubation period

2    distribution, $f_\tau$ of length $\tau$ days, in the following analysis.

3                                                    ---[Fig. 1 inserted about here]---



5    **Statistical analysis**

6        By definition, the serial interval $s$ is decomposed as the sum of time from the onset

7    of a primary case to secondary transmission, $u$, and incubation period of secondary

8    cases, $f$, i.e.,

9                                              $$s = u + f$$                              (1)

10    Thus, supposing that the number of secondary transmissions occurring $u$ days after the

11    onset of primary case (including the days for $u < 0$) was $\lambda(u)$, the expected number of

12    serial intervals of length $t$ is given by the convolution equation:

13                  $$\mathrm{E}(s_t) = \sum_i \sum_{k(i)} \sum_{u=-x}^{t} \lambda_u f_{t_{k(i)} - (t_i + u)}$$                  (2)

14    where $k(i)$ denotes the secondary cases infected by primary case $i$, and their times of

15    onsets are $t_{k(i)}$ and $t_i$, respectively. Considering that primary cases could acquire

16    infectiousness before the onset of disease, I assume the potentially contagious period to

17    $x = 2$ days before onset of primary case $i$. Two days was selected because more than

18    90% of observed cases experienced an incubation period longer than 2 days (i.e., the



1    secondary transmission from primary to secondary cases cannot occur before the

2    primary case gets infected). Since $s_t$ and $f_{\tau-u}$ are given, $\lambda(u)$, the frequency of secondary

3    transmissions relative to disease-age, can be determined using the deconvolution

4    procedure [18]. It should be noted that this approach assumes an independence of

5    secondary transmissions during the course of disease. Moreover, various intrinsic (i.e.

6    varying contact frequencies by disease-age due to disease progression) and extrinsic

7    factors (e.g., isolation measures) influencing the frequency of transmissions are ignored.

8        Since both $s_t$ and $f_{\tau-u}$ are known, estimates of the disease-age specific frequency of

9    secondary transmissions can be obtained in a non-parametric fashion [19], assuming a

10   step function model for $\lambda(u)$. Owing to the limited number of observations, the number

11   of parameters is restricted to six:

12
$$\begin{aligned}
\lambda(u) &= \rho_1 \quad for \ -2 \le u < 0 \\
\lambda(u) &= \rho_2 \quad for \ 0 \le u < 1 \\
\lambda(u) &= \rho_3 \quad for \ 1 \le u < 2 \\
\lambda(u) &= \rho_4 \quad for \ 2 \le u < 3 \\
\lambda(u) &= \rho_5 \quad for \ 3 \le u < 4 \\
\lambda(u) &= \rho_6 \quad for \ 4 \le u < 19 \\
\lambda(u) &= 0 \quad otherwise.
\end{aligned}$$
(3)

13   Referring to the observed maximum serial interval, I assume that Day 18 is the

14   maximum disease-age to observe secondary transmission. Assuming that $\lambda(u)$ is



1    generated by a nonhomogeneous Poisson process, resulting in $s_t$ serial intervals of

2    length $t$, the likelihood function, which is needed to estimate $\lambda(u)$, is proportional to

3    $$\prod_{t=-2}^{\infty} \left( \mathrm{E}(s_t) \right)^{r_t} \exp\left( -\mathrm{E}(s_t) \right)$$    (4)

4    where $r_t$ denote the daily counts of the serial interval. The maximum likelihood

5    estimates of parameters ($\rho_u$) that constitute $\lambda(u)$ were obtained by minimizing the

6    negative logarithm of equation (4). The 95% confidence intervals (CI) were determined

7    using the profile likelihood.

8        An extension was then made to account for the lethal course of illness in pneumonic

9    plague. Figure 1b shows the disease-age specific survivorship of primary pneumonic

10   plague (n = 166; i.e., the survival curve of the time from onset to death [11]). The mean

11   (median and SD) was 2.3 (2.0, 1.7) and the maximum length of survival was 12 days. It

12   should be noted that the probability of survival for 4 days after onset was only 4.8%

13   (95% CI: 0.8, 8.1), reflecting extremely acute and severe course of illness [2]. Although

14   the above simple model reasonably suggests the relative frequency of secondary

15   transmissions, the estimate is most likely biased by survivorship of cases. Thus,

16   regarding the obtained frequency of secondary transmissions as an implication of the

17   disease-age specific infectiousness, adjustment for the underestimation of the

18   infectiousness during later stage of disease is important. Let $g(u)$ the disease-age



1   specific probability of survival which is assumed to be known (as shown in Figure 1b), I

2   replaced the step function model (equation (3)) by the following:

3
$$\lambda(u) = \omega_1 g(u) \quad for \ -2 \leq u < 1$$
$$\lambda(u) = \omega_2 g(u) \quad for \ 1 \leq u < 2$$
$$\lambda(u) = \omega_3 g(u) \quad for \ 2 \leq u < 3$$
$$\lambda(u) = \omega_4 g(u) \quad for \ 3 \leq u < 6 \tag{5}$$
$$\lambda(u) = \omega_5 g(u) \quad for \ 6 \leq u < 9$$
$$\lambda(u) = \omega_6 g(u) \quad for \ 9 \leq u < 12$$
$$\lambda(u) = 0 \quad otherwise.$$

4   The model assumes the maximum disease-age to cause secondary transmission is Day

5   11 according to survivorship. Intervals for the piecewise constants differ from equation

6   (3) to appropriately capture the observed patterns of secondary transmission. The

7   likelihood function was derived in the same way (i.e., equation (4)) and the maximum

8   likelihood estimates of parameters ($\omega_u$) were obtained.



10   **RESULTS**

11       Figure 2a shows the estimated daily frequency of secondary transmissions with

12   corresponding 95% CI. During the second day of disease (i.e., 24–48 hours after onset),

13   the model estimated that 31.0% (95% CI: 11.6, 50.4) of the total number of secondary

14   transmissions had occurred. During the third day, the daily frequency of secondary

15   transmissions was second highest, yielding an estimate of 28.0% (10.2, 45.8). The



1  expected cumulative frequency of secondary transmissions at the end of third day was

2  81.9 %. Figure 2b compares observed and expected serial intervals. The $\chi^2$ goodness-of-

3  fit test revealed no significant deviation between these frequencies ($\chi^2_3 = 3.15$, p $= 0.37$;

4  see legend for Figure 2). The estimated $\lambda(u)$ suggests that the secondary transmissions

5  occurred immediately after, or before, onset of disease, indicating that countermeasures

6  before onset of disease such as chemoprophylaxis and quarantine should be

7  implemented to effectively control the transmissions.

8                                                      ---[Fig. 2 inserted about here]---

9       Figure 2c shows the survivorship-adjusted daily frequency of secondary

10  transmissions, $\omega_u$, by disease-age $u$. The expected frequency exhibits a bimodal pattern

11  with peaks at third (15.0% (95% CI: 9.5, 20.4)) and 10–12th (12.5% (95% CI: 4.9,

12  20.2)) days. From Day 4 of disease, the adjusted frequency of secondary transmissions

13  increased according to disease-age, suggesting that 76.3% of secondary transmissions

14  occurred during this period. Figure 2d compares observed and expected serial intervals,

15  confirming good agreement of the model with data ($\chi^2_3 = 0.53$, p $= 0.91$). It is difficult

16  to explain the bimodal pattern of disease-age specific infectiousness by biological

17  mechanisms only (i.e., natural history), and the low frequency between Days 3–6 may

18  not indicate that the cases in these disease-ages are less infectious than other stages.



1     Thus, the frequency is still likely biased by both intrinsic and extrinsic factors to deem

2     as disease-age specific infectiousness. Both the numbers of those who are considered as

3     infectious and those hospitalized and isolated widely vary over the course of illness [8].

4     Therefore, the extended model can only suggest that the infectiousness in later stage of

5     illness (i.e., disease-age greater than 4 days) is not insignificant.



## DISCUSSION

8     The present study investigated the frequency of secondary transmissions of primary

9     pneumonic plague relative to disease-age. The simplest model suggested that more than

10     four-fifths of the secondary transmissions occurred before the end of the third day,

11     supporting the need to implement countermeasures during the pre-clinical period or

12     immediately after onset. Public health measures such as quarantine, chemoprophylaxis

13     and isolation should be instituted as early as possible to effectively control the

14     transmissions.  The extended model incorporated the survivorship function, the mean of

15     which was as small as 2.3 days. Despite wide uncertainty of the frequency in later stage

16     of disease, the survivorship-adjusted frequency of secondary transmissions implied that

17     the infectiousness in the later stage of illness is non-negligible suggesting that it is

18     inappropriate to consider infectiousness in later stage is insignificant only from the



1    simple model. The adjusted frequency was still thought to be biased by underlying

2    factors such as isolation. Thus, more explicit clarification requires additional

3    information with regard to contact frequency and interventions. Despite the simplistic

4    assumptions, the present study is the first to explicitly estimate the disease-age specific

5    frequency of pneumonic plague based on empirical evidence.

6       Since this method would offer less costly and reasonable evaluation of disease-age

7    specific contagiousness and secondary transmissions and because this approach has the

8    potential to quantify the effectiveness of isolation measures, further evaluation with

9    different datasets are planned. In conclusion, this study applied the backcalculation

10    method to estimate the frequency of secondary transmissions, showing that known

11    distributions of the incubation period and serial interval offer key information relevant

12    to disease-control.





## REFERENCES


1. Inglesby TV, Dennis DT, Henderson DA, Bartlett JG, Ascher MS, Eitzen E, *et al*. Plague as a biological weapon: Medical and public health management. Working Group on Civilian Biodefense. *JAMA*. 2000; 283(17): 2281-2290.

2. Wu LT. A Treatise on Pneumonic Plague. Geneva: League of Nations Health Organization; 1926.

3. Park S, Chan KS, Viljugrein H, Nekrassova L, Suleimenov B, Ageyev VS, *et al*. Statistical analysis of the dynamics of antibody loss to a disease-causing agent: plague in natural populations of great gerbils as an example. *J R Soc Interface*. 2007; 4(12): 57-64.

4. Gani R, Leach S. Epidemiologic determinants for modeling pneumonic plague outbreaks. *Emerg Infect Dis*. 2004; 10(4): 608-614.

5. Massin L, Legrand J, Valleron AJ, Flahault A. Modelling outbreak control for pneumonic plague. *Epidemiol Infect*. 2007; 135(5): 733-739.

6. Fraser C, Riley S, Anderson RM, Ferguson NM. Factors that make an infectious disease outbreak controllable. *Proc Natl Acad Sci USA*. 2004; 101(16): 6146-6151.

7. Anderson RM, May RM. *Infectious Diseases of Humans*: *Dynamics and Control*. Oxford: Oxford University Press, 1991.





1    8. Nishiura H, Eichner M. Infectiousness of smallpox relative to disease age: estimates

2        based on transmission network and incubation period. *Epidemiol Infect.* 2007;

3        135(7):1145-50.

4    9. Fine PE. The interval between successive cases of an infectious disease. *Am J*

5        *Epidemiol.* 2003; 158(11): 1039-1047.

6    10. Wallinga J, Lipsitch M. How generation intervals shape the relationship between

7        growth rates and reproductive numbers. *Proc R Soc B.* 2007; 274(1609): 599-604.

8    11. Nishiura H. Epidemiology of a primary pneumonic plague in Kantoshu, Manchuria,

9        from 1910 to 1911: statistical analysis of individual records collected by the

10        Japanese Empire. *Int J Epidemiol.* 2006; 35(4): 1059-1065.

11    12. Nishiura H, Schwehm M, Kakehashi M, Eichner M. Transmission potential of

12        primary pneumonic plague: time inhomogeneous evaluation based on historical

13        documents of the transmission network. *J Epidemiol Community Health.* 2006;

14        60(7): 640-645.

15    13. Temporary Quarantine Section, Kanto Totokufu. *Epidemic record of plague during*

16        *1910-1911 (Meiji 43,4-nen 'Pest' Ryuko-shi).* Dalian: Manchurian Daily Press;

17        1912 (in Japanese).





1    14. Tieh TH, Landauer E, Miyagawa F, Kobayashi G, Okayasu G. Primary pneumonic

2        plague in Mukden, 1946, and report of 39 cases with 3 recoveries. *J Infect Dis.*

3        1947; 82(1): 52-58.

4    15. Brygoo ER, Gonon M. Une epidemie de peste pulmonaire dans le nord-est de

5        Madagascar. *Bull Soc Pathol Exot Filiales.* 1958; **51**: 47-60 (in French).

6    16. Ratsitorahina M, Chanteau S, Rahalison L, Ratsifasoamanana L, Boisier P.

7        Epidemiological and diagnostic aspects of the outbreak of pneumonic plague in

8        Madagascar. *Lancet.* 2000; 355(9198): 111-113.

9    17. Kasai K. A summary of the clinical symptoms observed in the plague pneumonia

10       that raged in south Manchuria during the months of January to March, 1911. PART

11       I. Presentation of evidence regarding the epidemic. In: *Report of the international*

12       *plague conference. Held at Mukden, April, 1911.* Manila: Bureau of Printing;

13       1912: 171-178.

14   18. Brookmeyer R, Gail MH. A method for obtaining short-term projections and lower

15       bounds on the size of the AIDS epidemic. *J Am Stat Assoc.* 1988; 83: 301-308.

16   19. Becker NG, Watson LF, Carlin JB. A method of non-parametric back-projection

17       and its application to AIDS data. *Stat Med.* 1991; 10: 1527-1542.






## FIGURE LEGENDS

**Figure 1. The incubation period distribution and probability of survival of primary pneumonic plague**

**(a)** The incubation period was observed during the largest outbreak in Manchuria from 1910–11 (n = 126, Data source: Kasai [17]). **(b)** Non-parametric probability of survival (solid line) with 95% confidence interval (broken line) of primary pneumonic plague (n = 166, Data source: Temporary Quarantine Section [13]).

**Figure 2. Frequency of secondary transmissions and predicted serial intervals**

**(a & c)** Expected daily frequency of secondary transmissions with corresponding 95% confidence intervals. Disease-age $t = 0$ denotes the onset of fever. **(b & d)** Observed and predicted daily counts of the serial intervals (n = 177, extracted from [13-16]). For both **(b)** and **(d)**, the $\chi^2$ tests revealed no significant deviations between the observed and predicted values ($\chi^2_3 = 3.15$, p = 0.37 and $\chi^2_3 = 0.53$, p = 0.91, respectively). When assessing the goodness-of-fit, serial intervals were divided into nine groups (<2, 2, 3, 4, 5, 6, 7, 8–10 and >10 days). **(a & b)** shows the results using simple model, and **(c & d)** shows the results of the extended model.



1    Figure 1

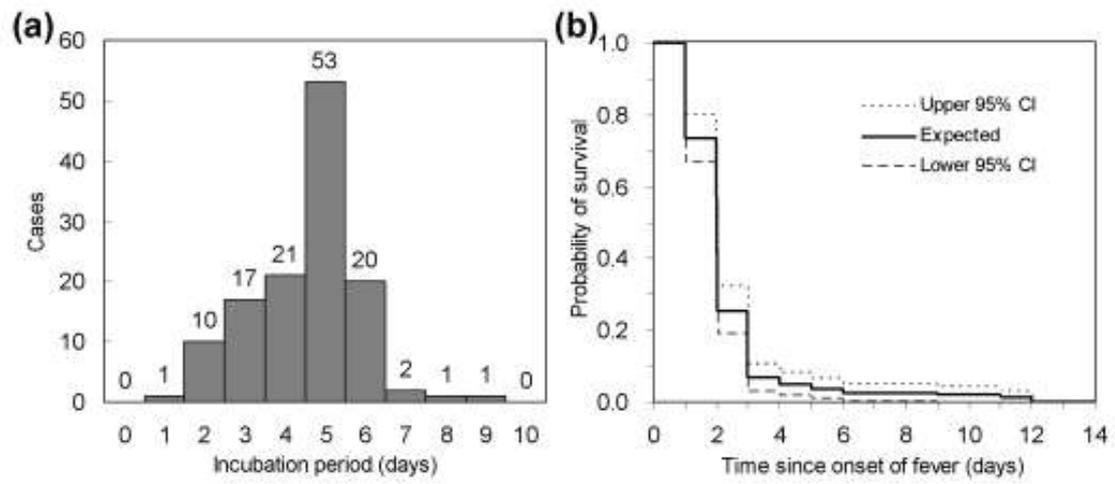





1    Figure 2

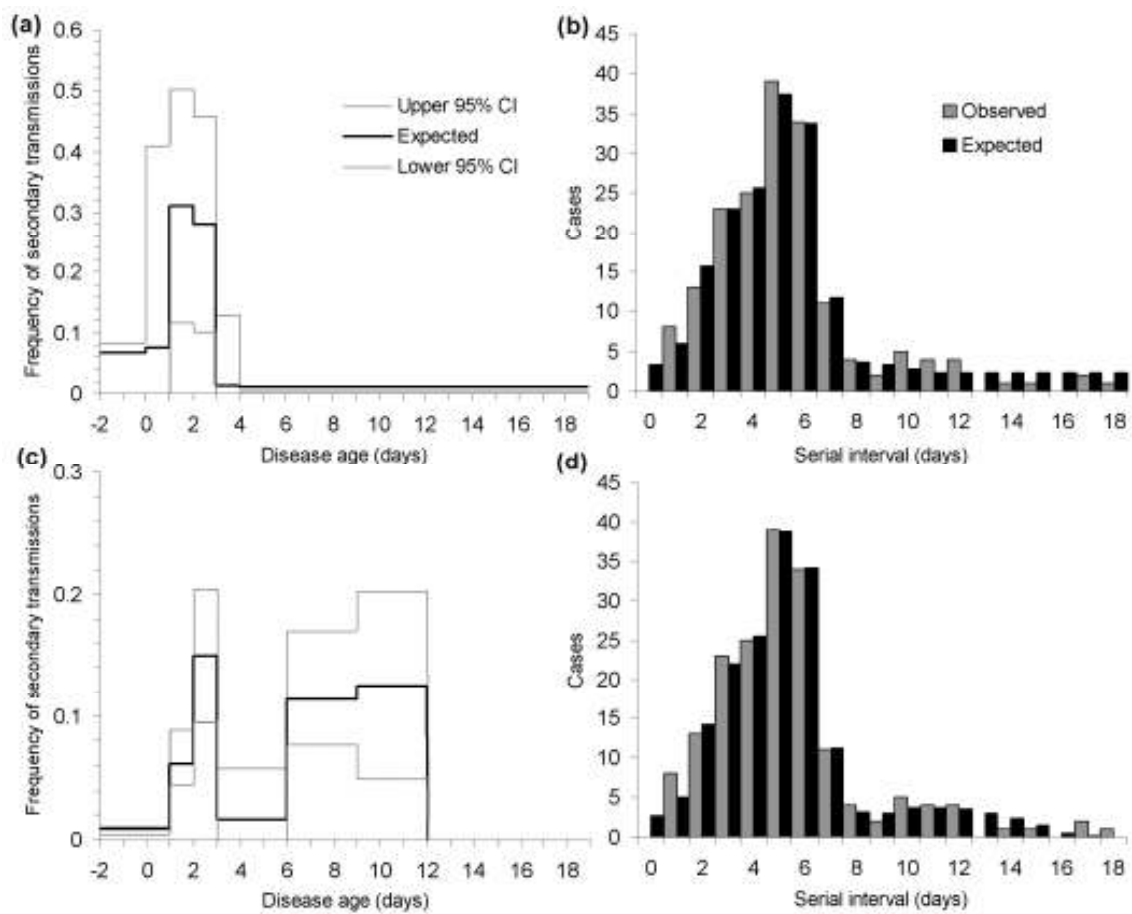